\documentclass[prl,twocolumn,showpacs,superscriptaddress,floatfix]{revtex4}
\usepackage{graphicx,amsfonts,amssymb,amsmath, hyperref}

\newif\ifhyper
\hypertrue
\ifhyper
\hypersetup{
   citecolor = {green},
   colorlinks = {true}, 
   urlcolor = {blue} 
}
\fi

\newcommand{\beq}{\begin{equation}}
\newcommand{\eeq}{\end{equation}}
\newcommand{\beqa}{\begin{eqnarray}}
\newcommand{\eeqa}{\end{eqnarray}}
\newcommand{\ket} [1] {\vert #1 \rangle}
\newcommand{\bra} [1] {\langle #1 \vert}

\begin{document} 

\title{First order phase transition in the anisotropic quantum orbital compass model}

\author{Rom\'an Or\'us}
\email{orus@physics.uq.edu.au}
\affiliation{The University of Queensland, School of Physical Sciences,  
QLD 4072, Australia}

\author{Andrew C. Doherty}
\email{doherty@physics.uq.edu.au}
\affiliation{The University of Queensland, School of Physical Sciences,  
QLD 4072, Australia}

\author{Guifr\'e Vidal}
\email{vidal@physics.uq.edu.au}
\affiliation{The University of Queensland, School of Physical Sciences,  
QLD 4072, Australia}

\begin{abstract}
We investigate the anisotropic quantum orbital compass model on an infinite square lattice by means of the infinite projected entangled-pair state algorithm. For varying values of the $J_x$ and $J_z$ coupling constants of the model, we approximate the ground state and evaluate quantities such as its expected energy and local order parameters.  We also compute adiabatic time evolutions of the ground state, and show that several ground states with different local properties coexist at $J_x = J_z$. All our calculations are fully consistent with a first order quantum phase transition at this point, thus corroborating previous numerical evidence. Our results also suggest that tensor network algorithms are particularly fitted to characterize first order quantum phase transitions. 

\end{abstract}
\maketitle

{\it Introduction.-} When quantum many-body systems are cooled down close to zero temperature, important collective phenomena may occur \cite{Schadev}. A good example is provided by transition-metal oxides, whose physical properties have become of increasing interest in the last few years \cite{review}. In these compounds the orbital degrees of freedom of the atomic electrons play a key role in determining properties such as metal-insulator transitions, high-temperature superconductivity and colossal magnetoresistance. 

The paradigmatic approach to these systems is based on the so-called orbital compass models \cite{OQC, classical}, which have been the subject of many studies in the past both in the classical and quantum regimes. For these systems, Jahn-Teller effects produce an anisotropy of the pseudospin couplings which is intertwined with the orientation of the interaction bonds. The properties of these systems have attracted considerable attention since they are endowed with symmetries that effectively reduce the dimensionality of the system (the so-called dimensional reduction) \cite{Dim, Dim2}. Despite of their apparent simplicity, orbital compass models are relevant in a variety of contexts, such as in determining the physics of Mott insulators with orbital degrees of freedom \cite{OQC} and the implementation of protected qubits for quantum computation in Josephson junction arrays \cite{pqubit}. 
These systems are also candidates to exhibit topological quantum order \cite{tqo}. Furthermore, it was recently shown how to simulate these models using polar molecules in optical lattices and systems of trapped ions with state-of-the-art technology \cite{optical, opticalLong}. 

Generally speaking, the symmetries in these systems involve large degeneracies in their energy spectra, which make their numerical simulation difficult \cite{finTquantum}. This fact, together with the lack of exact solutions, makes it hard to elucidate their phase diagrams. In this paper we use a tensor product state (TPS) \cite{TPS4, TPS5} or projected entangled-pair state (PEPS) \cite{PEPS} to study the \emph{two-dimensional anisotropic quantum orbital compass model (AQOCM)} and, in particular, to investigate whether its phase transition is of first order \cite{Ex, Mean} or second order \cite{XuMo}. More specifically, we use the infinite PEPS (iPEPS) algorithm of Ref. \cite{iPEPS} to study the model directly in the \emph{thermodynamic limit}. Our results provide abundant evidence in favor of a first order phase transition.

{\it The model.-} The 2D AQOCM describes a system of spins $1/2$ interacting on a square lattice with anisotropic two-body interactions as defined by the Hamiltonian
\beq
H = - J_x \sum_{\langle i,j \rangle} X^{[i,j]} X^{[i+1,j]} -J_z \sum_{\langle i,j \rangle} Z^{[i,j]} Z^{[i,j+1]} \ ,
\label{ham}
\eeq
where $X^{[i,j]}$ ($Z^{[i,j]}$) is the Pauli X (Z) operator at site $[i,j]$ of the lattice, and $J_x$ ($J_z$) is the coupling in the $x$ ($z$) direction. 

For this model, Nussinov and Fradkin \cite{Sym} proved that its Hamiltonian is dual to a plaquette model proposed by Xu and Moore to describe $p+ip$ superconducting arrays such as Sr$_2$RuO$_4$ \cite{XuMo}. The influence of impurities \cite{Imp} and of diluted lattices \cite{dilution} in the model has also been investigated. In addition, finite temperature properties have been studied both in the quantum and classical versions of the model \cite{finTquantum, finTclass}, and in both cases the existence of a low temperature ordered phase with a thermal transition lying in the 2D Ising universality class has been shown. Finally, in Ref. \cite{1DCompass} a 1D version of the model was shown to undergo a first order phase transition.

The Hamiltonian from Eq. (\ref{ham}) has also some significant properties in the context of quantum computation. For instance, the model was proven to be dual to the 2D cluster state Hamiltonian embedded in a magnetic field \cite{Clus}. It was also shown to be related to certain classes of quantum error correcting codes where the system is used to codify a qubit that is robust against external local noise \cite{bacon}.

Before proceeding any further, let us sketch some of the basic symmetry properties of the Hamiltonian $H$ in Eq. (\ref{ham}) (see e.g. Refs \cite{Sym, bacon} for detailed discussions). Define the operators
\beq
P_i \equiv \prod_j X^{[i,j]} ~~~~~~~~~ Q_j \equiv \prod_i Z^{[i,j]} \ ,
\label{symmetry}
\eeq
where $P_i$ acts on column $i$ of the 2D lattice and $Q_j$ acts on row $j$. It is not difficult to check that these operators commute with $H$ for all the values of $i$ and $j$. Importantly, $[P_i,Q_j] \ne 0$ for any $i$, $j$, and therefore operators $P_i$ and $Q_j$ represent \emph{incompatible} symmetries of $H$. Furthermore, notice that $[P_i,P_{i'}] = 0 ~ \forall i, i'$ and similarly for $Q_j$, and that any tensor product of operators corresponding to different columns (or rows) commutes with $H$ as well. All these symmetries imply that, in the case of a system defined on an $L \times L$ square lattice, every eigenstate of the Hamiltonian is at least of order $O(2^L)$ degenerate. Also, whenever $J_x = J_z$ the system is invariant under the reflection symmetry $X \leftrightarrow Z$, indicating the self-duality of the model at equal couplings \cite{Sym}. 

The above self-duality indicates a possible phase transition in the system at  $J_x = J_z$. There have been several attempts to determine the existence and order of this phase transition. On the one hand, Xu and Moore pointed towards a possible second order quantum phase transition \cite{XuMo}. On the other hand, some approximate calculations seem to favor a first order transition \cite{Ex, Mean}. The nature of this phase transition is, therefore, not totally understood yet.

\begin{figure}
 \includegraphics[width=0.48\textwidth]{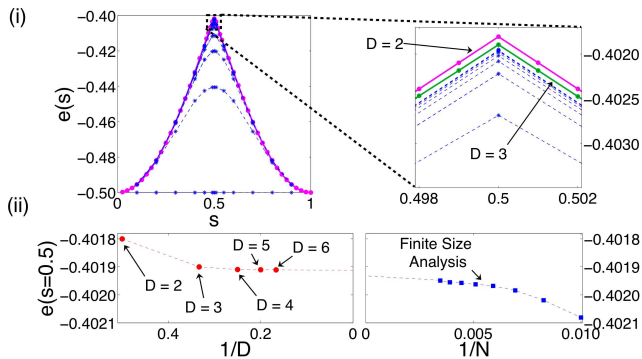}
 \caption{(color online) (i) Energy per link \emph{e} in the AQOCM on an infinite square lattice obtained by using the iPEPS algorithm with $D = 2,3$ (results for $D=4,5,6$ are very similar to those for $D=3$). The energy $e$ has a sharp peak with discontinuous derivative at the phase transition. Dotted lines correspond to the results from Ref. \cite{Ex} up to 16$\times$16 lattices using exact diagonalization and Green's function Montecarlo (plotted with permission). Lines linking numerical points are a guide to the eye. (ii) Comparison at $s = 0.5$ of the energy per bond computed with the iPEPS algorithm for $D = 2,\ldots,6$ and the finite-size analysis from Ref. \cite{Ex}. Lines linking numerical points are a guide to the eye.}
 \label{fig2}
 \end{figure}

{\it The method.-} 
In this paper we use the iPEPS algorithm \cite{iPEPS} to compute the ground state as well as adiabatic time evolutions for the AQOCM on an infinite 2D square lattice. As explained in Refs. \cite{PEPS, iPEPS}, the accuracy of the results relies on a refinement parameter that we shall refer to as $D$. This parameter is related to the maximum entanglement content that can be handled by the simulations \footnote{The entanglement entropy of a block of contiguous sites and with boundary of size $L$ is upper bounded in a PEPS by $L \log D$. Thus, PEPS can handle the 'boundary-law scaling' of the entanglement entropy for 2D systems \cite{PEPS}.}. In practice, increasing the value of $D$ leads to better descriptions of the ground state, and therefore to more accurate estimations of the different observable quantities. In our calculations we consider $D = 2,\ldots,6$ and, without loss of generality, $J_x, J_z \ge 0$ \footnote{It is always possible to make a local transformation that brings the Hamiltonian into this form and leaves the spectrum unchanged.}. The coupling strengths in Eq. (\ref{ham}) can be restricted to the range $J_x, J_z \in [0,1]$ and written in terms of a variable $s \in [ 0, 1]$ as $J_x = \cos{(s\pi/2)}$ and $J_z = \sin{(s\pi/2)}$. 


Let us discuss the impact that the symmetries of the system have in our simulations. As explained above, the symmetries of the AQOCM imply an infinite degeneracy of its ground state in the thermodynamic limit \cite{Sym}. For instance, different ground state wave functions can be labelled according to the different eigenvalues of operators $P_i$ in Eq. (\ref{symmetry}). This sort of degeneracy, however, does not play a significant role in our simulations since our representation of the quantum state by means of an iPEPS is, by construction, invariant under translations in the $x$ and $z$ directions \cite{iPEPS}. Still, our implementation of the algorithm could be sensitive to the two-fold degeneracy caused by a simultaneous flip of all the spins. In practice, however, we observe that this does not happen. The simulations spontaneously choose either a positive or negative value of $\langle X^{[i,j]} \rangle$ (or $\langle Z^{[i,j]} \rangle$) for all sites $[i,j]$ away from the phase transition point \footnote{Such a symmetry breaking can be understood as resulting from the finite value of $D$, which favors the collapse of the ground state into one of two low-entangled (symmetry-breaking) options, and not a superposition thereof.}. 

{\it Simulation results.-} Our calculations are of two types. First, we have computed the ground state wave function $\ket{\Psi_{GS}(s)}$ of the system as a function of $s$ and evaluated observable quantities on it such as energy and local order parameters. Second,  we have simulated adiabatic time evolutions starting from the computed ground state $\ket{\Psi_{GS}(s_{ini})}$ for a given initial parameter $s_{ini}$, and adiabatically increasing or decreasing $s$ in the Hamiltonian well beyond crossing the point $s = 1/2$ ($J_x = J_z$). These evolutions define two families of states, the left $\ket{L(s)}$ for $s_{ini}  < 1/2$, and the right $\ket{R(s)}$ for $s_{ini}  > 1/2$. 
 
\begin{figure}
 \includegraphics[width=0.48\textwidth]{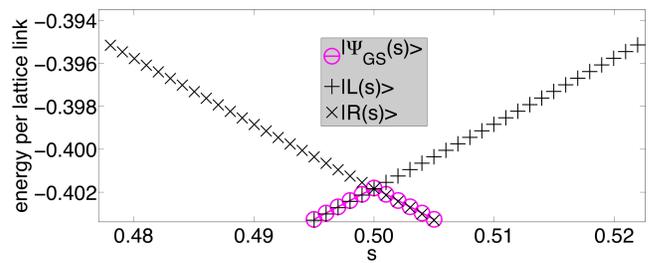}
 \caption{(color online) Expected energy per lattice link for the ground state $\ket{\Psi_{GS}(s)}$ and the adiabatically evolved states $\ket{L (s)}$ and $\ket{R (s)}$, as computed with the iPEPS algorithm with $D = 2$.}
 \label{figAdiEn}
 \end{figure}

The ground state energy per lattice link
\begin{eqnarray}
	e(s)&\equiv& \frac{J_x}{2} \bra{\Psi_{GS}(s)} X^{[i,j]} X^{[i+1,j]}\ket{\Psi_{GS}(s)} \nonumber \\
	 &+& \frac{J_z}{2} \bra{\Psi_{GS}(s)}Z^{[i,j]} Z^{[i+1,j]} \ket{\Psi_{GS}(s)},
\end{eqnarray}
(independent of $i$ and $j$) is displayed in Fig. (\ref{fig2}). Our results show the presence of a sharp peak at $s =1/2$, which is compatible with the existence of a first order phase transition at this point. The energy per link in the adiabatically evolved states $\ket{L(s)}$ and  $\ket{R(s)}$ is also plotted in Fig. (\ref{figAdiEn}). There we can see that the energy of e.g. $\ket{L(s)}$ follows the ground state energy up to the transition point $s = 1/2$. More generally, we find that, up to numerical accuracy, the PEPS for $\ket{L(s)}$ is the same as that for the ground state for $s < 1/2$ (and similarly for $\ket{R(s)}$ in the regime $s > 1/2$). Therefore,  
\begin{equation*} 
\ket{\Psi_{GS}(s)} \sim \left\{ 
\begin{array}{rl} 
\ket{L(s)}  & \text{if } s < 1/2\\ 
\\
\ket{R(s)}  & \text{if } s > 1/2
\end{array} \right. \ . 
\label{gs}
\end{equation*} 
From Fig. (\ref{figAdiEn}) we can also infer that state $\ket{L(s)}$ no longer corresponds to the ground state of the system for $s > 1/2$, but rather to some higher-energy excitation (and similarly for $\ket{R(s)}$ for $s < 1/2$). The simulations of $\ket{L(s)}$ and $\ket{R(s)}$ are robust against modifying the rate of change of the Hamiltonian during the adiabatic evolution, indicating the presence of an energy gap to the reachable excitations. At the phase transition point both states  $\ket{L(1/2)}$ and $\ket{R(1/2)}$ have the same energy as the actual ground state $\ket{\Psi_{GS}(1/2)}$, indicating the presence of two possible ground states of the system at this point.

\begin{figure}[h]
 \includegraphics[width=0.48\textwidth]{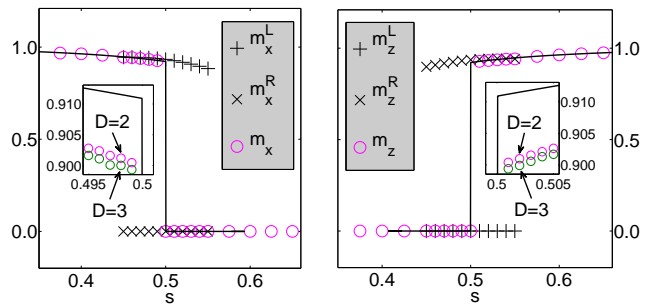}
 \caption{(color online) Expected values of the local order parameter operators $X^{[i,j]]}$ and $Z^{[i,j]}$ (in absolute value) for the ground state  $\ket{\Psi_{GS}(s)}$ ($m_x$ and $m_z$) and the adiabatically evolved states  $\ket{L (s)}$ ($m_x^L$ and $m_z^L$) and $\ket{R (s)}$ ($m_x^R$ and $m_z^R$), obtained by using the iPEPS algorithm with $D = 2,3$ (results for $D=4,5$ are very similar to those for $D=3$).  The lines correspond to the results from Ref. \cite{Mean} using mean field theory after fermionization of the Hamiltonian (plotted with the author's permission).}
 \label{fig1}
 \end{figure}

 
Importantly, these two ground states at s = 1/2 can be shown to be locally different, for instance by computing the Ising-like order parameters
\begin{eqnarray}
	m_x(s) &\equiv& \bra{\Psi_{GS}(s)} X^{[i,j]} \ket{\Psi_{GS}(s)},	\\
	m_z(s) &\equiv& \bra{\Psi_{GS}(s)} Z^{[i,j]} \ket{\Psi_{GS}(s)},
\end{eqnarray}
which are independent of $[i,j]$ due to translation invariance.  Fig. (\ref{fig1}) shows $m_x$ and $m_z$ as a function of $s$, together with analogous expected values $m_x^{L}(s)$, $m_z^{L}(s)$, $m_x^{R}(s)$ and $m_z^{R}(s)$ for the evolved states $\ket{L(s)}$ and $\ket{R(s)}$. We find that $m_x$ and $m_z$ are both discontinuous at $s = 1/2$. However, such discontinuity could originate in a lack of resolution in $s$. That is, perhaps by considering more points around $s=1/2$, the discontinuity in the order parameters would disappear, indicating a continuous phase transition. This possibility can be ruled out by noticing that e.g. $m_x^{L}(s)$ does not vanish to the right of the transition point (similarly, $m_z^{R}(s)$ does not vanish to the left of the transition point). That is, the two families of states $\ket{L(s)}$ and $\ket{R(s)}$, which coincide with the ground state to the left (respectively right) of $s=1/2$, remain locally different at the transition point, where both represent possible ground states of the system. We interpret this fact as conclusive evidence of the existence in the 2D AQOCM of a first order phase transition between the two phases characterized by vanishing and non-vanishing values of the local order parameters $m_x$ and $m_z$. 
 
Let us now discuss the role played by the symmetries in this phase transition. Our numerical calculations using tensor networks have also shown that the ground states $\ket{\Psi_{GS}(s)}$ satisfy the eigenvalue relations $P_i \ket{\Psi_{GS}(s)} = \ket{\Psi_{GS}(s)}$ if $s < 1/2$ and $Q_j \ket{\Psi_{GS}(s)} = \ket{\Psi_{GS}(s)}$ if  $s > 1/2$, 
regardless of the values of $i$ and $j$. Thus, we see that the system chooses to preserve a different symmetry at each side of the phase transition point, namely, the $P_i$ symmetry for $s < 1/2$ and the $Q_j$ symmetry for $s > 1/2$. Quite naturally, the system chooses to break the symmetry which minimizes the amount of entanglement in the broken ground state, while leaving the remaining symmetry intact. In turn, this also implies that the adiabatically evolved states  $\ket{L(s)}$ and  $\ket{R(s)}$ are, respectively, eigenstates of operators $P_i$ and $Q_j$ with eigenvalue $1$ for \emph{any} value of $s$. This follows from the fact that the symmetry of the initial state is preserved all along the adiabatic time evolution since the symmetry operators commute with the Hamiltonian for any value of $s$. Therefore, the two possible ground states at the phase transition point $\ket{L(1/2)}$ and $\ket{R(1/2)}$ obtained by adiabatic evolution preserve the $P_i$ and $Q_j$ symmetries respectively. 
 
In addition, we observe that the two families of adiabatically evolved states are related to each other by a non-local transformation, namely the duality transformation of the model that switches the values of $J_x$ and $J_z$ in  Eq. (\ref{ham}). More precisely, for all the computed values of $s$, these are related by a rotation $\ket{L(s)} = W(\pi/2) w(\pi/2) \ket{R(s)}$,  where $W(\pi/2)$ rotates the spin degrees of freedom by an angle $\pi/2$ around the $y$-axis and $w(\pi/2)$ rotates the square lattice by $\pi/2$. That it takes a highly non-local transformation to map $\ket{L(s)}$ and $\ket{R(s)}$ into each other is, again, consistent with a first order transition, where the two coexisting ground states are not expected to be connected by local perturbations.

Furthermore, we have also computed the ground state fidelity-per-site diagram \cite{fid1,fid11, fid2} for this system (not shown) and have obtained results that agree with the typical behavior expected of a first order transition (see Ref. \cite{fid2}).

All the above results are compatible with those obtained using other numerical approaches. As a first check, we have verified that our simulations reproduce the results of simple series expansion calculations that we performed far away from $s = 1/2$. 
As can be seen in Fig. (\ref{fig2}), the present results for the energy per bond $e$, computed directly for an infinite system, agree in the first 4 significant digits with the value obtained through a rough extrapolation, to the thermodynamic limit, of exact diagonalization and Green's function Montecarlo results for finite systems presented in Ref. \cite{Ex}. Moreover, as shown in Fig. (\ref{fig1}), close to the phase transition point the present results for the order parameters $m_x$ and $m_z$ are comparable to those obtained in Ref. \cite{Mean} with mean field theory after fermionization of the Hamiltonian. The small disagreement, of the order of 1.5 $\%$, increases with growing values of $D$ (that is, as our results become more precise), which suggests that the iPEPS results for $D=2$ are already better than those obtained by combining fermionization with mean field theory. We stress that our simulations show fast convergence of the computed observables with the refinement parameter $D$ (see e.g. Fig. (\ref{fig2}.(ii))). 
 


{\it Conclusions.-} In this paper we have provided fresh evidence that, contrary to what had been suggested in Ref. \cite{XuMo}, the phase transition in the AQOCM on a square lattice is of first order. Unlike previous approaches to this problem, we have employed an algorithm based on a TPS or PEPS for an infinite 2D lattice to numerically compute the ground state and, for the first time for an infinite 2D system, its adiabatic time evolution. We believe that our results, together with those in Ref. \cite{Ex, Mean}, \emph{conclusively} support the existence of a first order phase transition. 


{\it Acknowledgements.-} The authors thank H.-D. Chen and J. Dorier and acknowledge financial support from The University of Queensland (ECR2007002059) and the Australian Research Council (FF0668731, DP0878830).


{}

\end{document}